\documentclass[10pt,twocolumn,twoside]{osajnl}
\journal{pr} 

\setboolean{shortarticle}{false}
\usepackage{amsmath}
\usepackage{amssymb}
\usepackage{graphicx}
\usepackage{dcolumn}
\usepackage{bm}
\usepackage{color}
\usepackage{caption}
\usepackage{subfigure}
\usepackage{float}
\usepackage{lineno}
\usepackage{appendix}
\usepackage{hyperref}
\usepackage{upgreek}
\usepackage{cuted}
\usepackage{amsmath,lipsum}
\usepackage{cuted}
\usepackage{multicol}
\usepackage{siunitx}

\title{Continuous-variable quantum key distribution over 50.4 km fiber using integrated silicon photonic transmitter and receiver}

\author[1,2†]{Shuaishuai Liu}
\author[1,2†]{Yanxiang Jia}
\author[1,2]{Yuqi Shi}
\author[1,2]{Yizhuo Hou}
\author[4]{Pu Wang}
\author[1,2]{Yu Zhang}
\author[1,2]{Shiwei Yang}
\author[1,2]{Zhenguo Lu}
\author[1,2,3*]{Xuyang Wang}
\author[1,2,3*]{Yongmin Li}
\affil[1]{State Key Laboratory of Quantum Optics Technologies and Devices, Institute of Opto-Electronics, Shanxi University, Taiyuan 030006, China}
\affil[2]{Collaborative Innovation Center of Extreme Optics, Shanxi University, Taiyuan 030006, China}
\affil[3]{Hefei National Laboratory, Hefei 230088, China}
\affil[4]{School of Information, Shanxi University of Finance and Economics, Taiyuan 030006, China}

\affil[*]{Corresponding author: wangxuyang@sxu.edu.cn, yongmin@sxu.edu.cn}

\begin{abstract}
Quantum key distribution (QKD) is the fastest-growing and relatively mature technology in the field of quantum information, enabling information-theoretically secure key distribution between two remote users. Although QKD based on off-the-shelf telecom components has been validated in both laboratory and field tests, its high cost and large volume remain major obstacles to large-scale deployment. Photonic integration, featured by its compact size and low cost, offers an effective approach to addressing the above challenges faced by QKD. Here, we implement a high-performance, integrated local local oscillator continuous-variable (CV) QKD system based on an integrated silicon photonic transmitter and receiver. By employing a high-speed silicon photonic integrated in-phase and quadrature modulator, a low-noise and high bandwidth silicon photonic integrated heterodyne detector, and digital signal processing, our CV-QKD system achieves a symbol rate of up to 1.5625 GBaud. Furthermore, the system achieves asymptotic secret key rates of 31.05 and 5.05 Mbps over 25.8 and 50.4 km standard single-mode fiber, respectively, using an 8-phase-shift keying discrete modulation. Our integrated CV-QKD system with high symbol rate and long transmission distance pays the way for the quantum secure communication network at metropolitan area.
\end{abstract} 

\setboolean{displaycopyright}{true}

\begin{document}

\maketitle

\section{Introduction}
The rapid development of quantum computing \cite{1,2,3} heralds a potential threat to classical encryption algorithms and will render traditional encryption methods based on the computational complexity vulnerable. Quantum key distribution (QKD), based on the principles of quantum mechanics, enables information-theoretically secure key exchange \cite{4,5,6,7,8,9,10}, and its security is guaranteed against adversaries even with unlimited computational power.\\
\indent Current research on QKD primarily focuses on three aspects: security, high performance, and practicality. In terms of security, researchers are dedicated to enhancing QKD’s resistance to Eve’s attacks. This includes security proofs against coherent attacks \cite{11,12}, measurement-device-independent QKD \cite{13,14,15,J.-Y. Liu}, source-device-independent QKD \cite{16}, one-sided-device-independent QKD \cite{17}, and device-independent QKD \cite{18}, all aimed at enhancing the security and robustness of QKD systems against Eve’s attacks on both the quantum channel and side channels. For high performance, researchers focus on improving device performance \cite{19}, enhancing system stability \cite{20,21}, increasing multiplexed channels \cite{22,23,Y. Wang}, boosting system repetition rate \cite{24,25,W. Li}, and modifying modulation schemes \cite{25,26,27} to enhance system performance and overcome distance limitations. The secret key rate of the repeaterless point-to-point QKD is limited by the Pirandola-Laurenza-Ottaviani-Banchi (PLOB) bound \cite{28}. Interestingly, twin-field (TF) QKD is capable of surpassing this limit \cite{29,30,31,32}, achieving long-distance transmission over 1000 km in optical fiber links \cite{30}. Discrete-variable QKD (DV-QKD) and continuous-variable QKD (CV-QKD) each has its own advantages. For intercity backbone networks spanning hundreds of kilometers, TF-QKD can be employed. In contrast, the quantum networks within metropolitan areas, where high-speed communication over distances of a few dozen kilometers is required, CV-QKD is particularly well-suited \cite{33,34,35}. CV-QKD can be constructed using standard telecom components \cite{J. Aldama}, including in-phase and quadrature (IQ) modulators for quantum state preparation and coherent detectors for quantum state measurement \cite{C. Bruynsteen,J. F. Tasker,F. Raffaelli}. In terms of practicality, the system's cost and volume are crucial factors determining whether QKD can be commercially deployed on a large scale. Traditional CV-QKD box system are costly and bulky. By utilizing photonic integrated circuit (PIC) technology to integrate the photonic components of the transmitter (receiver) on a single platform, the system's cost and size can be dramatically reduced \cite{36,37,J. Wang,A. E.-J. Lim}, which attracted widespread attention in recent years.\\
\indent At the same symbol rate, traditional Gaussian-modulated CV-QKD offers advantages in secret key rate \cite{P.Wang}, transmission distance \cite{20,21}, and security proofs \cite{38}. However, the system requires high-resolution digital-to-analog converters (DACs) to handle continuous Gaussian-modulated signals, which impose great challenges for high-speed modulation formats above GBaud level \cite{25}. Discrete modulation can effectively address this issue by using a finite, discrete constellation, making it compatible with high-speed systems. The performance of the traditional quadrature phase shift keying (QPSK) (four-state) discrete modulation is significantly worse than that of Gaussian-modulation protocols, which diminishes the advantages of discrete modulation. To improve the performance of the discrete modulation, medium scale constellations are employed. More precisely, amplitude and phase shift keying (APSK), along with quadrature amplitude modulation (QAM) with constellations of 64, 128, and 256 have been demonstrated \cite{25,26,40,F. Roumestan}. Later, it was found that a two ring constellation with only a dozen of states can achieve performance very close to the Gaussian modulation \cite{27,P.Wang}.\\
\indent Recently, a proof of principle experiment of low speed (0.8 MBaud) Gaussian-modulation CV-QKD with integrated silicon photonics was demonstrated over a simulated channel \cite{41}. Additionally, several CV-QKD systems with integrated receiver chips \cite{25,42,43,44,Y. Jia} have been reported. These achievements promote the progress of on-chip integrated CV-QKD to realize an integrated GBaud rate CV-QKD system, not only the receiver, but also the transmitter should be integrated on a PIC. On the other hand, GBaud symbol rate requires that both the integrated transmitter and receiver can operate over bandwidth of several GigaHertzs (GHz), as well as low-noise, high fidelity modulation (transmitter) and shot-noise-limited detection (receiver). At present, an integrated GBaud rate CV-QKD system over long-distance single-mode fiber where both the transmitter and receiver integrated on PICs has not been reported, to our knowledge.\\
\indent In this article, we report a integrated high-speed CV-QKD system using a high bandwidth silicon photonic transmitter and receiver over a long-distance fiber channel. The modulation signals at the transmitter are shaped by a pre-emphasis filter, and the output signals from the receiver are equalized by a digital post-equalizer. The combination of the two approaches effectively suppressing the modulation noise and measurement noise in the system, and enable a high fidelity encoding and decoding of quantum signals at symbol rate of 1.5625 GBaud. By using eight symbol phase-shift keying (PSK) modulation and well-designed digital signal processing (DSP), our integrated QKD system achieves the asymptotic secret key rates of 31.05 and 5.05 Mbps over 25.8 and 50.4 km standard single mode fibers, respectively.\\
\begin{figure}[ht] 
	\centering\includegraphics[width=7.5cm]{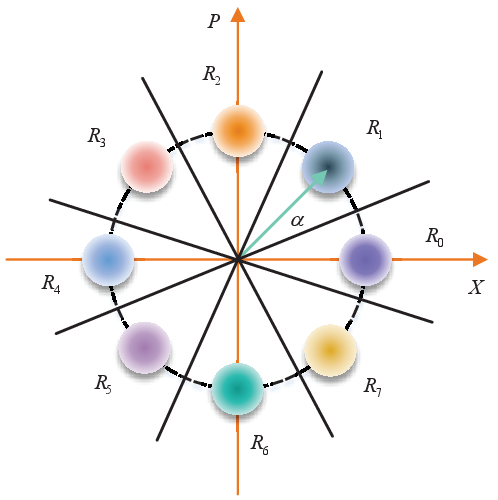}\\
	\caption{\label{The 8-state constellation diagram} The constellation diagram of 8-PSK. The solid circles of different colors represent eight quantum states. $R_j$ represents the key mapping region of Bob where he maps his measurement results to the discretized key values.}
\end{figure} 
\section{Discrete-modulation CV-QKD protocol}
\subsection{The protocol description}
\subsubsection{State preparation}
\begin{figure}[ht] 
	\centering\includegraphics[width=8.8cm]{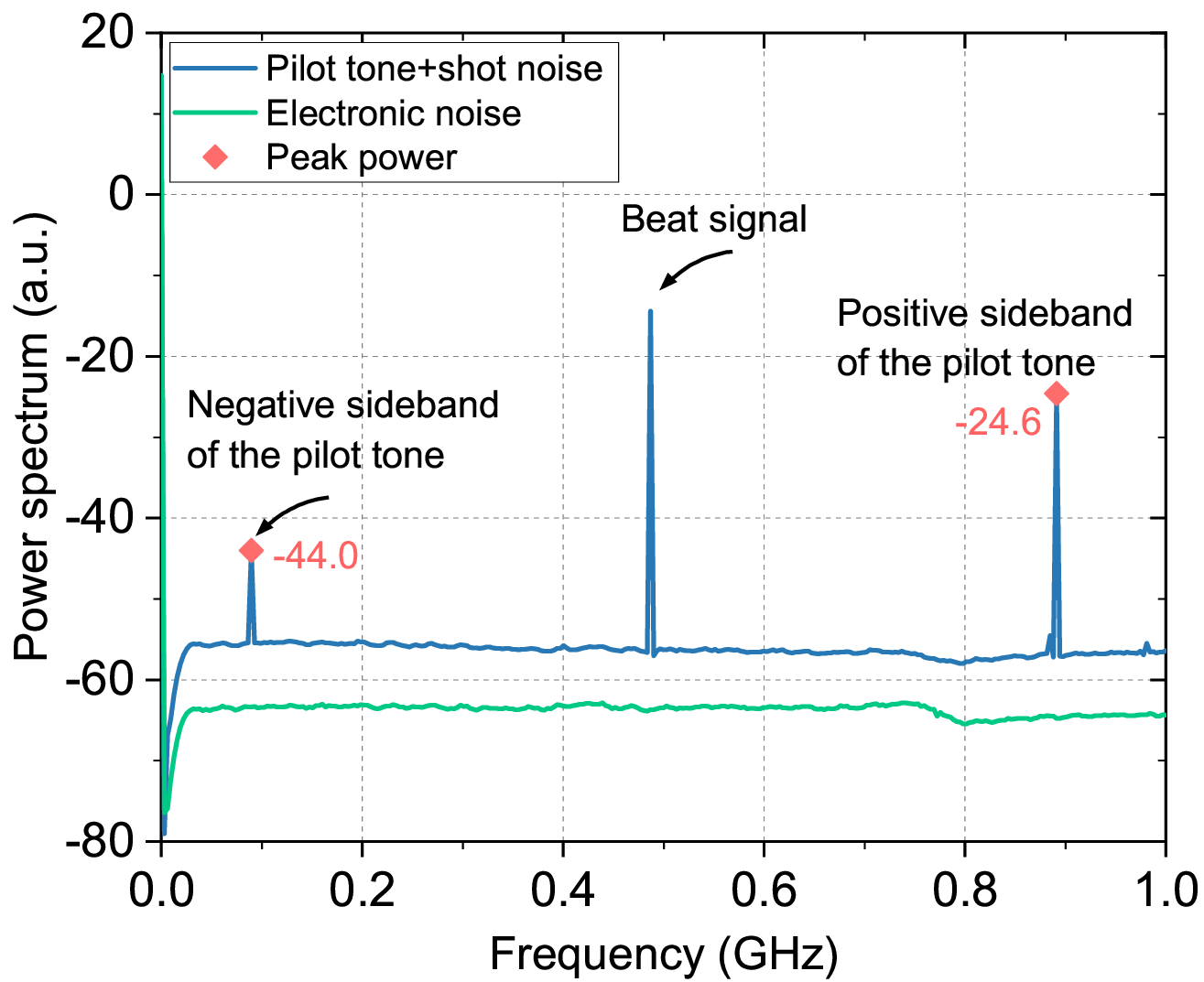}\\
	\caption{\label{The sideband suppression ratio} Sideband suppression ratio. The red diamond represents the peak power of the positive and negative sidebands of the pilot tone.}
\end{figure}
Our system uses an 8-PSK discrete modulation CV-QKD protocol to distribute keys, achieving a relatively high secret key rate with a small, simple constellations. Fig. \ref{The 8-state constellation diagram} shows a schematic of the constellation of 8-PSK. The 8 quantum states are randomly and uniformly prepared by Alice in the phase space, where each state having an equal preparation probability of 1/8 and is represented as $\left| {{\alpha }_{k}} \right\rangle =\left| \alpha \exp \left( ik\pi /4 \right) \right\rangle ,k\in \left\{ 0,1,2,3,4,5,6,7 \right\}$. The modulation variance of the quantum signal is ${{V}_{\text{A}}}=2{{\alpha }^{2}}$, which is normalized to the shot noise units (SNUs). \\
\indent Carrier suppressed single-sideband modulation (CS-SSB) is a type of sideband modulation in which the carrier and one of the sidebands are suppressed. To realize CS-SSB by an IQ modulator that consists of a pair of Mach-Zehnder modulators (MZMs) nested in a Mach-Zehnder interferometer (MZI), both MZMs operate at the null bias point to suppress the optical carrier and the relative phase between the I and Q paths is set at 90° (-90°) to eliminate negative (positive) first-order sideband. The output field of the IQ modulator is given by
\begin{equation}\label{1}
\begin{aligned}
	& {{E}_{\text{out}}}\left( t \right)=\frac{1}{2}{{E}_{\text{in}}}\left( t \right)(\sin \left[ \mu \left( t \right)\cos \left( {{\omega }_{\text{s}}}t+\theta \left( t \right) \right) \right] \\ 
	&\quad \quad \quad\quad\; +\sin \left[ \mu \left( t \right)\sin \left( {{\omega }_{\text{s}}}t+\theta \left( t \right) \right) \right]\exp \left[ i\pi \frac{{{V}_{\text{PM}}}}{2{{V}_{\pi /2\;}}} \right]), \\ 
\end{aligned}
\end{equation}
where
\begin{equation}\label{2}
\begin{aligned}
&{{E}_{\text{in}}}\left( t \right)={{E}_{0}}\exp \left( i{{\omega }_{0}}t \right),\quad \mu \left( t \right)=\frac{r\left( t \right)\cdot \pi }{2{{V}_{\pi }}}, \\ &r(t)=\sqrt{{{K}_{\text{I}}}{{(t)}^{2}}+{{K}_{\text{Q}}}{{(t)}^{2}}},\\
\end{aligned}
\end{equation}
${{E}_{\text{in}}}$ denotes the input field, ${{K}_{\text{I}}}(t)$ and ${{K}_{\text{Q}}}(t)$ are the real and imaginary parts of the complex baseband quantum signal, respectively, $\theta \left( t \right)$ is the phase of the baseband quantum signal, ${{V}_{\pi }}$ is the half-wave voltage, ${{V}_{\text{PM}}}={{V}_{\pi /2}}$ is the bias voltage between the I and Q paths, ${{\omega }_{\text{s}}}$ denotes the frequency shift applied to the quantum signal. For weak modulation depths $\mu \left( t \right)\ll 1$, Eq. (\ref{1}) can be simplified as \cite{N. Jain}:
\begin{equation}\label{3}
{{E}_{\text{out}}}\left( t \right)\approx {{E}_{0}}{{J}_{1}}[\mu \left( t \right)]\text{exp}i[\left( {{\omega }_{0}}+{{\omega }_{\text{s}}} \right)t+\theta \left( t \right)],
\end{equation}
where ${{J}_{1}}$ is the Bessel function of the first kind of orders 1. Eq. (\ref{3}) indicates that an ideal CS-SSB modulation can be achieved with a perfect IQ modulator.\\
\indent For a realistic CS-SSB modulation (assuming that a positive first-order sideband is modulated), the negative first-order sideband (image sideband) cannot be completely suppressed. In our experiment, the on-chip IQ modulator has a suppression ratio of negative sideband exceeding 19.4 dB (Fig. \ref{The sideband suppression ratio}). Considering that the negative first-order sideband is the dominant residual term under the weak modulation condition, the ratio of the positive sideband amplitude to the total sideband amplitude can be given by \cite{33}:
\begin{equation}\label{4}
	d\approx \sqrt{\frac{\sigma }{\mu +\sigma }}=\sqrt{\frac{1}{1.0115}}\approx 0.9943,
\end{equation}
where $\mu$ and $\sigma$ denote the power of the positive (negative) first-order sideband. To defend against the potential attacks due to the negative first-order sideband leakage, we correct Alice's data of quantum signal to incorporate the leaked information into the prepared quantum state
\begin{equation}\label{5}
	{{\alpha }^{'}}=\alpha /d.
\end{equation}
In this case, the leaked information through negative first-order sideband is attributed to the insecure quantum channel, and a reliable secure key rate can be obtained.
\subsubsection{System calibration}
In the system calibration, the system parameters for determining the secret key rate are calibrated. Firstly, the detection efficiency $\eta $ and electronic noise ${{v}_{\text{el}}}$ of the heterodyne detection are measured. Then, the modulation variance ${{V}_{\text{A}}}$ of the quantum signal is calibrated by Alice and Bob with a back to back configuration, where the transmitter and receiver are directly connected by using a short single mode fiber. In this case, the channel transmittance is known with $T=1$.\\
\subsubsection{Distribution, measurement and parameter estimation}
\indent The calibrated quantum signal is sent through the insecure quantum channel to Bob, who measures the received quantum and pilot signals using the heterodyne detection. Based on the pilot tone, Bob can recover the frequency and phase of the quantum signal and obtain $\left( {{x}_{\text{B}}},{{p}_{\text{B}}} \right)$. For parameter estimation, Alice and Bob randomly disclose a small part of their data. Bob calculates the first-moment $\left( \left\langle {{x}_{\text{B}}} \right\rangle ,\left\langle {{p}_{\text{B}}} \right\rangle  \right)$ and second-moment $\left( \left\langle x_{\text{B}}^{2} \right\rangle ,\left\langle p_{\text{B}}^{2} \right\rangle  \right)$ observables of the quadrature ${{x}_{\text{B}}}$ and ${{p}_{\text{B}}}$, respectively. Then he estimates the secret key rate by using numerical convex optimization techniques \cite{P.Wang,46}. \\ 
\indent The channel transmittance $T$ and excess noise $\varepsilon$ are also estimated
\begin{equation}\label{6}
		T=2\frac{{{\langle {{x}_{\text{A}}}{{x}_{\text{B}}}\rangle }^{2}}}{{\eta }{{\langle x_{\text{A}}^{2}\rangle }^{2}}},
\end{equation}
\begin{equation}\label{7}
		\varepsilon =\frac{\left\langle x_{\text{B}}^{2} \right\rangle -1-{{\upsilon }_{\text{el}}}}{{\eta T}/{2}\;}-{{V}_{\text{A}}},
\end{equation}
 where ${{x}_{\text{A}}}$ represents Alice’s modulation data. Notice that, the channel transmittance and excess noise only play a secondary role in the security and performance estimation for the discrete modulation protocol.\\
\subsubsection{Post-processing: Key map, error correction and privacy amplification}
The raw key string of Bob is obtained via the following key map. Exactly, each quantum signal measured by Bob is labeled as 
$y=\left| y \right|\exp \left( i\theta  \right)$, according to the region $R_{j}$
\begin{equation}\label{8}
	z= j,\text{if}~\theta \in \left[ \frac{(2j-1)\pi }{8},\frac{(2j+1)\pi }{8} \right)\to y\in {{R}_{j}}, 
\end{equation}
where $j\in \left\{ 0,1,2,3,4,5,6,7 \right\}$, the raw key $(k,z)$ is obtained.\\
\indent Subsequently, error correction is performed to obtain identical bit strings, and the potentially leaked information is removed through privacy amplification, resulting in the generation of the final secure key.\\
\begin{figure}[!ht] 
	\centering\includegraphics[width=8.8cm]{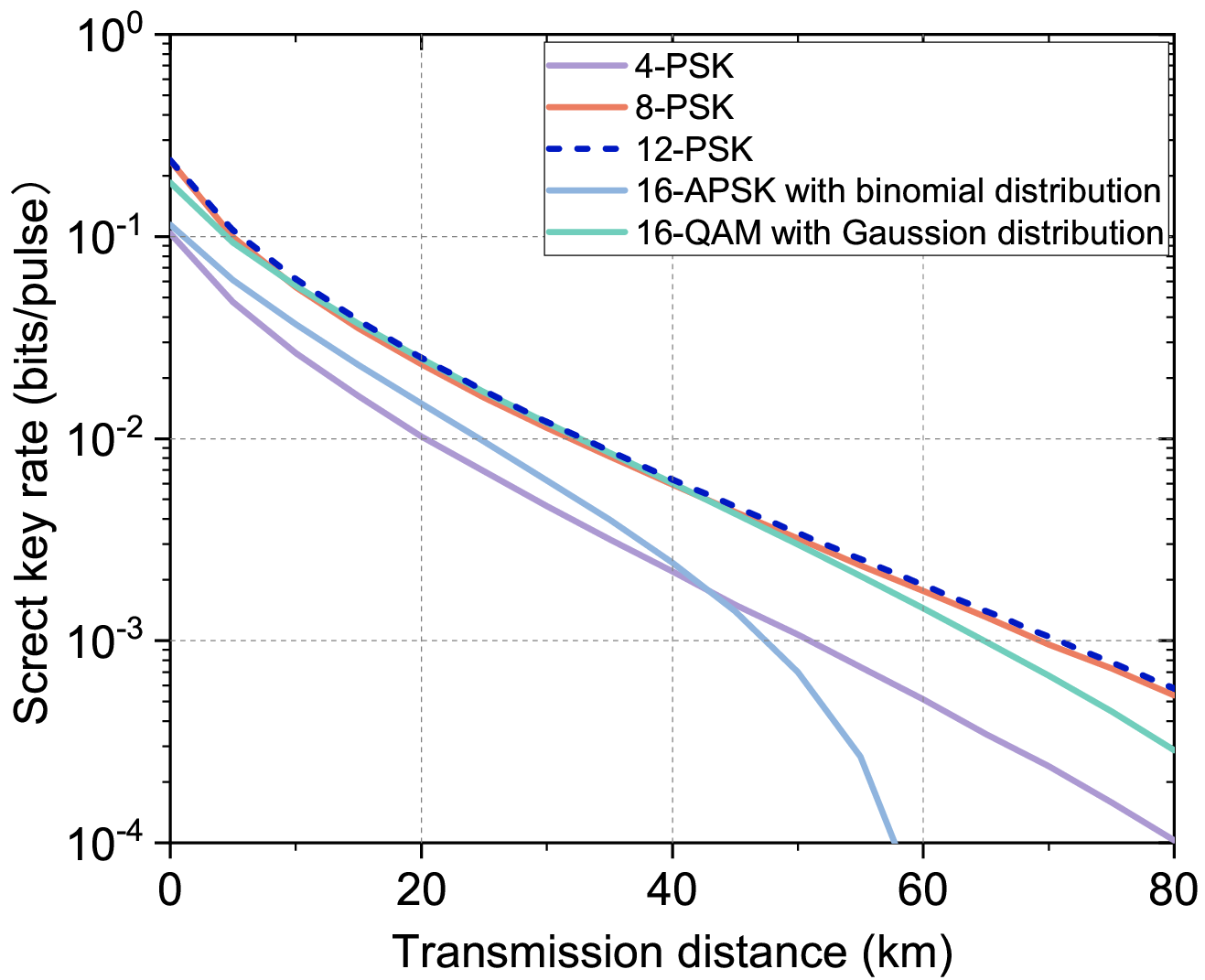}\\
	\caption{\label{different modulation protocols and transmission distances} 
		The secret key rate of different discrete modulation protocols versus the transmission distances. The modulation variance have been optimized. Other parameters are set to reverse reconciliation efficiency $\beta =0.95$, excess noise $\varepsilon=0.031 $, detection efficiency $\eta=0.37 $, and electronic noise for heterodyne detection ${{v}_{\text {el}}}= 0.20$.}
\end{figure}\\
\subsection{Performance analysis}
In Fig. \ref{different modulation protocols and transmission distances}, we plot the secret key rate of different discrete modulation protocols versus the transmission distances (a channel loss of 0.2 dB per kilometer is assumed). For different protocols, the key rate has been optimized by adopting the optimal modulation variance at each distance and the rest of the parameters are the same. In comparison to the traditional 4-PSK modulation, 8-PSK achieves over a 132\% increase in secret key rate at 25 km and the improvement becomes more significant as the transmission distance increases. At present, there are two kinds of security proof approaches for the discrete modulation CV-QKD protocols, the numerical convex optimization technique \cite{45}, and analytical methods \cite{A. Denys}. In comparison with the analytical method, the numerical optimization approach gives a tighter bound (higher secret key rate) at the price of being more computationally intensive. Here the PSK protocol utilizes the numerical convex optimization approach to establish the security proof, whereas 16-APSK and 16-QAM rely on analytical methods for security proof. However, further replacing the 8-PSK with 12-PSK does not significantly improve the key rate. The performance of the 16-APSK and 16-QAM under the security frame of the analytical secret key rate is comparable to the 4-PSK and 8-PSK, respectively \cite{A. Denys}. Therefore, in our experiment we choose the 8-PSK scheme with numerical optimization approach to achieve the good trade-off between the performance and modulation complexity of the protocol.
\begin{figure*}[!ht] 
	\centering\includegraphics[width=18cm]{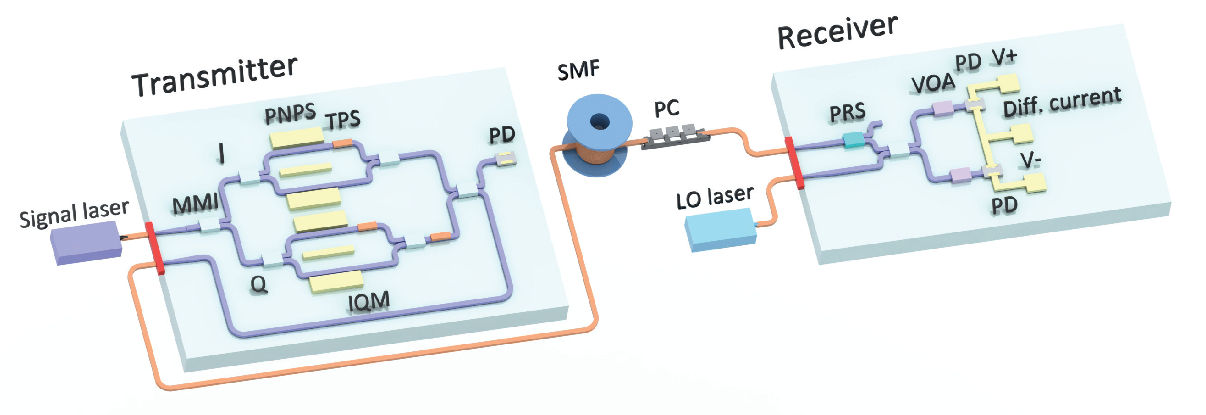}\\
	\caption{\label{CV-QKD system} Schematic of the silicon photonic integrated CV-QKD system. MMI, multimode interferometer; IQM, in-phase and quadrature modulator; PNPS, p-n junction phase shifter; TPS, thermal phase shifter; PD, Photodiode; PC, polarization controller; PRS, polarization rotation splitter; VOA, variable optical attenuator.}
\end{figure*} 
\subsection{Rate}
In the asymptotic case, the secret key rate of the 8-PSK discrete modulation CV-QKD against collective attacks is given by \cite{P.Wang,45,46,S. Ghorai}
\begin{equation}\label{10}
	{{K}^{\infty }}= \underset{{{\rho }_{\text{AB}}}\in \mathcal{S}}{\mathop{\min }}\,D\left( \mathcal{G}\left( {{\rho }_{\text{AB}}} \right)\|\mathcal{Z}\left[ \mathcal{G}\left( {{\rho }_{\text{AB}}} \right) \right] \right)-{{p}_{\text{pass}}}{{\delta }_{\text{EC}}},
\end{equation}
where $D\left( \rho \left\| \sigma  \right. \right)=\text{Tr}\left( \rho \text{lo}{{\text{g}}_{2}}\rho  \right)\text{-Tr}\left( \rho \text{lo}{{\text{g}}_{2}}\sigma  \right)$ is the quantum relative entropy of the operators $\rho$ and $\sigma $; $\mathcal{G}$ is positive and trace non-increasing map for post-processing processes. ${{\rho }_{\text{AB}}}$ is the joint state of the Alice and Bob after Alice’s quantum state has been transmitted through a quantum channel. The set $S$ includes all the density operators that are consistent with the experimental observations. $\mathcal{Z}$ represents a pinching quantum channel for achieving key mapping outcomes. ${{p}_{\text {pass}}}$ stands the probability of selecting raw data, in our experiment ${{p}_{\text {pass}}}$ equal to 1. ${{\delta }_{EC}}$ is the information exposed during error correction process and can be given by: 
\begin{equation}\label{11}
	\begin{aligned}
		& {{\delta }_{\text{EC}}}=H\left( Z \right)-\beta I\left( X;Z \right) \\ 
		& \ \ \quad \text{=}\left( 1-\beta  \right)H\left( Z \right)\text{+}\beta H\left( Z\left| X \right. \right),\text{ } \\ 
	\end{aligned}
\end{equation}
where $X$ and $Z$ are the raw data of Alice and Bob, respectively, $H (Z)$ denotes the Shannon entropy, $\beta$ is the reverse reconciliation efficiency, $I (X;Z)$ is the Shannon mutual information between Alice and Bob, and $H (Z|X)$ is the conditional entropy.
\section{Integrated transmitter and receiver}
\begin{figure*}[!ht] 
	\centering\includegraphics[width=18cm]{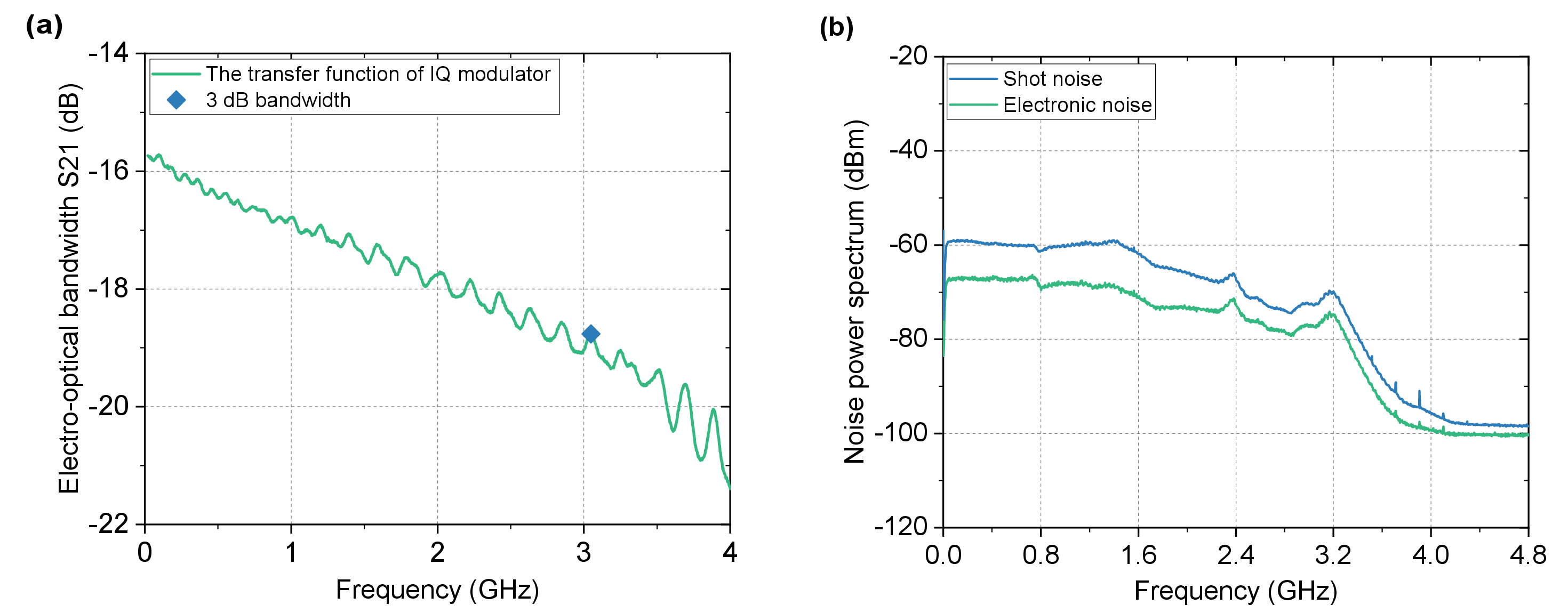}\\
	\caption{\label{The spectrum diagrams} Frequency response of the IQ modulator and the heterodyne detector. (a) Frequency response of the IQ modulator. The 3 dB bandwidth of the IQ modulator 3.05 GHz. (b) Frequency response of the heterodyne detector. The shot-noise-limited bandwidth of the heterodyne detector exceeds 3.2 GHz.}
\end{figure*}

Figure ~\ref{CV-QKD system} shows the schematic of our silicon photonic integrated CV-QKD system. The transmitter uses an external tunable laser and the input and output beams are coupled into and from the chip through edge coupler array (ECA). The ECA comprise of a polarization maintaining (PM) fiber array with high numerical aperture (NA = 0.4) that are packaged on the edge of the chip. In the transmitter, the input beam transmits in transverse electric (TE) mode. The integrated IQ modulator consists of a pair of MZMs nested in a MZI. For high-speed radio frequency (RF) driving, each MZM uses phase shifters employing lateral p-n junction that works in the carrier depletion mode, and traveling-wave electrodes with impedance matching $ R = 50 \;\Omega$ are designed. The modulator adopts a series push-pull driving configuration with $V_{\pi}$ of 4 V. We measured the electro-optical transfer function (S21) of the MZMs via network analyzer, the Q branch and the I branch exhibit similar response. The results for the Q branch are depicted in Fig. \ref{The spectrum diagrams}(a) and the 3 dB modulation bandwidth electro-optical of 3.05 GHz was observed. The monitoring photodiode after the IQ modulator combined with a bias controller circuits module are employed to stabilize the bias points of the IQ modulator in real-time. The feedback signals are applied to the thermal phase shifters inside the MZMs and MZI for CS-SSB modulation. The transmitter exhibits an overall loss of 12.2 dB, of which 3.8 dB is attributed to the total edge-coupling insertion loss.\\
\indent In the receiver, the signal and local oscillator (LO) beams are coupled into the chip through ECA with high NA single mode fiber and PM fiber respectively (Fig. \ref{CV-QKD system}). The input polarization state of the signal beam is converted to TE-mode by using the polarization controller and a polarization rotation splitter. Subsequently, the signal and LO beams are directed to a heterodyne detector, which is composed of a 2$\times$2 50:50 multimode interferometer (MMI) and two Ge-Si photodiodes with a responsivity of 0.9A/W. To balance the intensities of two output paths of the MMI coupler, two variable optical attenuators based on p–i–n phase modulators are used, which introduces controllable losses owing to the free carrier absorption effect when forward voltage is applied. The insertion loss is 0.1 dB when no voltage is applied. The overall efficiency of the receiver was measured to be $\eta = 0.37$, with the majority of the loss attributed to the edger coupler insertion loss of 2.1 dB and the imperfect quantum efficiency of the photodiodes 0.72. Fig. \ref{The spectrum diagrams}(b) shows the noise power spectrum of the on-chip heterodyne detector with and without the LO beam, from the results we can determine that the shot-noise-limited bandwidth of the heterodyne detector exceeds 3.2 GHz. The common-mode rejection ratio (CMRR) of the integrated heterodyne detector is measured and over 23 dB CMRR is observed.\\
\indent For integrated receivers, on-chip photodiodes typically provide bandwidths exceeding 20 GHz, so the photodiode itself is not the primary limiting factor in achieving shot-noise-limited performance at high bandwidths. To achieve the shot noise limit under high-bandwidth, the design of the amplification circuit is critical. We use low noise figure, high gain ABA-52563 RF amplifier and small packages components in the amplifier circuit. Since the RF amplifier provides stable output gain without requiring a feedback network, which effectively reduces the parasitic capacitance, resistance, and inductance. Small packages components not only reduce parasitic capacitance and inductance, but also shorten the signal transmission path, thereby suppressing the transmission losses and enhancing the impedance matching. To shield the electromagnetic interference, we enclose the circuits within an aluminum shielding box. Above measures improve the detector’s high-frequency performance.
\begin{figure}[!ht] 
	\centering\includegraphics[width=8.8cm]{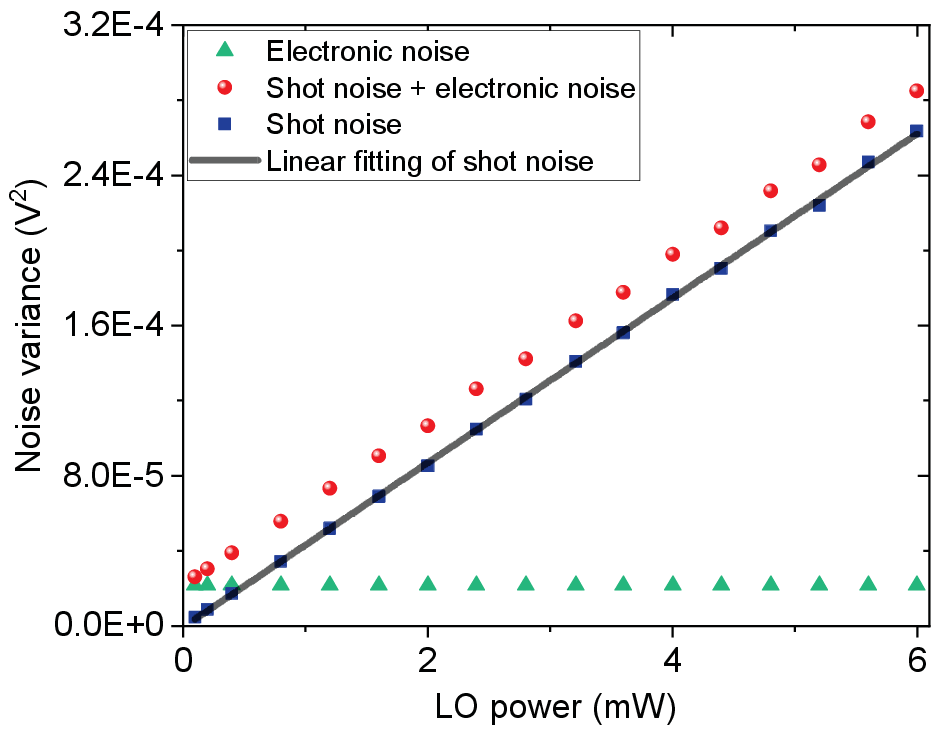}\\
	\caption{\label{The linear relationship} The linear dependence relationship between the power of the LO and the shot noise for the on-chip heterodyne detector. The green triangles, red circles, and blue squares represent the variances of electronic noise, overall noise, and shot noise, respectively. The black line indicates the linear fitting of the shot noise. }
\end{figure}
\begin{figure*}[!ht] 
	\centering\includegraphics[width=18cm]{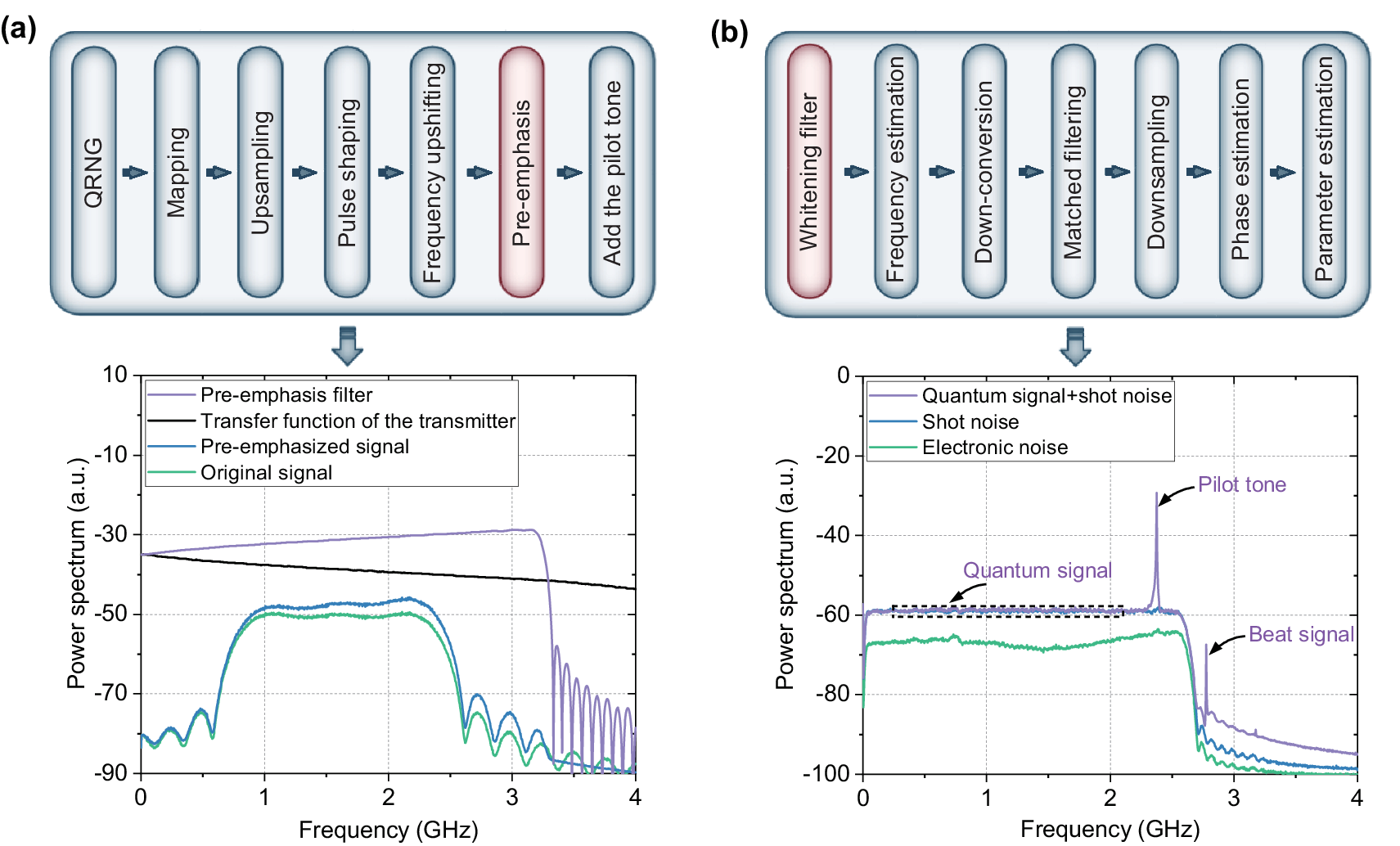}\\
	\caption{\label{DSP} The DSP of the integrated CV-QKD. (a) Preparation of the quantum signals and pilot tones. Alice generates the 8-PSK quantum signals in phase space using quantum random numbers generator and upsamples it to 25 GSample/s. The upsampled quantum signals are pulse-shaped by a digital RRC filter. The shaped signals are frequency-shifted to 1.6 GHz and compensated for the high-frequency root-roll-off of the IQ modulator using a pre-emphasis filter. Finally, a 400 MHz pilot signal is added. (b) Decoding of the quantum signals. Firstly, the signals from the heterodyne detection are compensated for the high-frequency attenuation of the heterodyne detector using a whitening filter. Subsequently, the pilot tone is transformed to the frequency domain using a fast Fourier transform (FFT), and the frequency difference between the quantum signal and the LO is estimated. Then, the quantum signal is frequency-shifted to baseband and filtered by a matched RRC filter to remove out-of-band noises. Finally, phase correction is applied to the quantum signals based on the phase estimation results from the pilot tone.}
\end{figure*} 
\section{Experimental implementation}
\subsection{Experimental System}
As shown in Fig.~\ref{CV-QKD system}, our experimental system consists of a silicon photonic integrated high-speed transmitter located at Alice's station, a quantum channel based on standard single-mode optical fiber, and a silicon photonic integrated high-bandwidth receiver located at Bob's station. A continuous single-frequency laser with a linewidth of 100 Hz at wavelength of 1550.12 nm (NKT Koheras BASIK X15) is used as the signal laser. Eight uniformly distributed quantum states in phase space are randomly prepared using a silicon photonic IQ modulator. To realize frequency and phase recovery and minimize the impact of the pilots on the quantum signal, the pilot signal is frequency multiplexed at a frequency 1.2 GHz away from the center frequency of the quantum signal. The IQ modulator operates in CS-SSB modulation mode, which suppress the influence of the carrier and the first negative sideband. A 10-bit resolution dual-channel arbitrary waveform generator (AWG) (Tektronix, AWG70002A) with the sampling rate of 25 GSample/s converts the generated digital signals into two RF analog signals to drive the IQ modulator. The system's symbol rate is set to 1.5625 GBaud. Finally, the quantum signal carrying the key information is coupled to the single-mode fiber via ECA and sent to Bob.\\
\indent At the Bob's station, he employs a laser of the same model with Alice’s laser to serve as the local LO (LLO), which has a frequency difference of 2.78 GHz in comparison to Alice’s laser. This configuration shifts the first negative sideband outside the detection bandwidth of the heterodyne detector, thereby minimizing its impact on the excess noise in the system. 
After the LO and the quantum signal coupled into the silicon photonic integrated receiver, they are mixed and interfered via the 2$\times$2 MMI. The interfered signals are fed into a pair of on-chip Ge-Si photodiodes and the generated photocurrents are subtracted and amplified by cascaded RF amplifiers. Figure ~\ref{The linear relationship} plots the output noise variance of the detector as a function of the LO power for a vacuum field input. The observed results confirm the linear response and shot-noise-limited characteristics of the detector. \\
\indent The electrical signals from heterodyne detectors are acquired by a 12-bit oscilloscope (MSO, Tektronix, MSO64B) at the sampling rate of 25 GSample/s. The acquired data and are stored on the computer’s hard drive and used for DSP. The overall process of the modulation, acquisition, and DSP \cite{48} are program-controlled and run automatically. 
\subsection{Digital Signal Processing}
\subsubsection{Transmitter}
A high-bandwidth silicon photonic IQ modulator is a prerequisite to realize a high Baud rate QKD. The bandwidth of an unpackaged bare-chip IQ modulator can reach above 10 GHz, due to chip packaging, the observed bandwidth of the IQ modulator is around 3.05 GHz (Fig. \ref{The spectrum diagrams}(a)). Additionally, the attenuation of the RF cables used in our experiment further results in the transmitter bandwidth to 1.23 GHz, as shown in Fig. \ref{DSP}(a). The limited bandwidth inevitably causes cross-talks between adjacent quantum signals for a gigahertz-level Baud rate \cite{25}, and not only disrupts the independent and identical distribution of the quantum signals that is a precondition in the security proof \cite{49}, but also increases the modulation noise, which further degrade the system’s performance.\\
\begin{figure*}[!ht] 
	\centering\includegraphics[width=18cm]{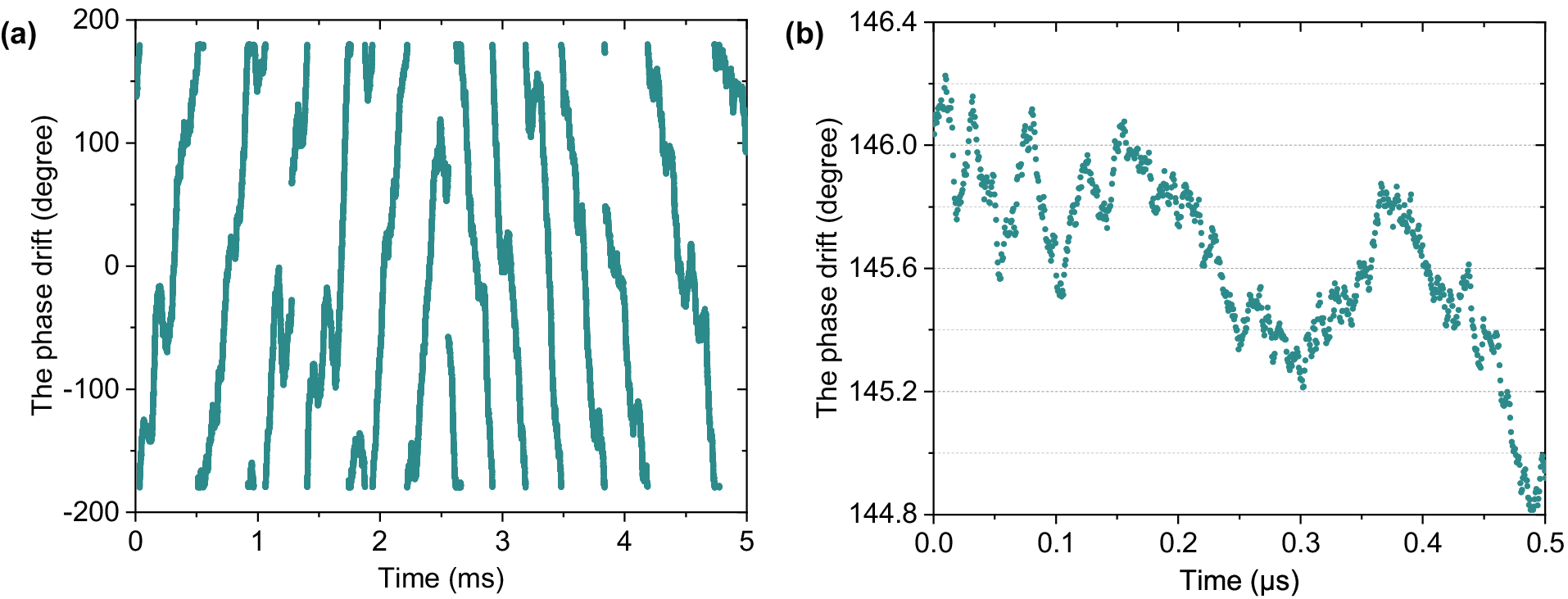}\\
	\caption{\label{The phase drift} Measured phase drift of the pilot tone. (a) The phase drift observed on the 5 ms timescale. (b) The phase drift within 0.5 $\SI{}{\micro\second}$.}
\end{figure*}
\indent To address this issue and ensure high-fidelity quantum state preparation, we designed a pre-emphasis filter to compensate for the high-frequency roll-off based on the measured transfer function of the transmitter. After the compensation, the cutoff frequency of the transmitter sets to 3.3 GHz, as shown in Fig \ref{DSP}(a). The detailed steps of the DSP process to generate the quantum signals and pilot tones are as follows: firstly, a series of uniformly distributed quantum random numbers is mapped into an 8-PSK constellation quantum signals with a rate of 1.5625 GBaud. Next, the quantum signals are upsampled to 25 GSample/s and pulse shaped through a digital root-raised-cosine (RRC) filter with a roll-off factor of 0.2. After this, the baseband quantum signal is up-shifted to 1.6 GHz. Then, a pre-emphasis digital filter based on the inverse frequency response of the transmitter is employed to pre-process the frequency shifting quantum signal. To generate the pilot tone, we directly generate cosine and sine waveforms at 400 MHz, sampled at 25 GSample/s. These waveforms are added to the frequency shifting quantum signal. Next, the composite signal is converted into an analog signal using an AWG, which is used to drive the on-chip IQ modulator to modulate the optical field. In this way, both the quantum signal and the pilot tone are simultaneously generated.\\
\begin{table*}[!ht]
	\caption{\label{tab:table} \textbf{Experimental parameters of our integrated CV-QKD system. $V_{\text{A}}$, modulation variance of the quantum signal;  ${{v}_{\text {el}}}$, electronic noise for heterodyne detection; $\eta $, overall detection efficiency; $\varepsilon $, excess noise (refers to the channel input); $T$, transmittance of the quantum channel; $\beta $, reconciliation efficiency; ${{K}^{\infty}}$, the asymptotic secret key rate.}}
	\centering
	\setlength{\tabcolsep}{2.3mm}{
		\begin{tabular}{ccccccccc}
			\hline
			\multicolumn{1}{c}{Distances (km)} & 
			\multicolumn{1}{c}{Symbol Rates (GBaud)} &
			\multicolumn{1}{c}{$V_{\text{A}}\left( \text{SNU} \right)$} & \multicolumn{1}{c}{${{v}_{\text {el}}}\left( \text{SNU} \right)$} & \multicolumn{1}{c}{$\eta \left( \text{ }\!\!\%\!\!\text{ } \right)$} & \multicolumn{1}{c}{$\varepsilon \left( \text{SNU} \right)$}& \multicolumn{1}{c}{$T$} & \multicolumn{1}{c}{$\beta$ (\%)}  & 
			\multicolumn{1}{c}{${{K}^{\infty}}$ (Mbps)} \\  \hline  
			25.8  & 1.5625& 1.19 & 0.20 & 37& 0.028 & 0.34 & 95  &  $31.05$  \\
			50.4  & 1.5625& 1.22 & 0.20 & 37& 0.033 & 0.11 & 95  &  $5.05$  \\ \hline
	\end{tabular}}
\end{table*}
\begin{table}[!ht]
	\caption{\label{tab:table2} \textbf{The first and second moments of the experimentally measured observable at 25 km. $QS_{m}$, the m$th$ quantum state, $m\in \left\{ 0,1,2,3,4,5,6,7 \right\}$; $ \left\langle {{x}_{\text{B}}} \right\rangle$ and $\left\langle {{p}_{\text{B}}} \right\rangle $, the first moment; $ \left\langle x_{\text{B}}^{2} \right\rangle$ and $\left\langle p_{\text{B}}^{2} \right\rangle $, the second moment.}}
	\centering
	\setlength{\tabcolsep}{3.6mm}{
		\begin{tabular}{c|cccc}
			\hline
			\multicolumn{1}{c|}{$QS_{0}$ $ \left\langle {{x}_{\text{B}}} \right\rangle ,\left\langle {{p}_{\text{B}}} \right\rangle $ }& 0.3859 & $3.62\times {{10}^{-4}}$   \\
			\multicolumn{1}{c|}{$QS_{0}$ $ \left\langle x_{\text{B}}^{2} \right\rangle ,\left\langle p_{\text{B}}^{2} \right\rangle  $ }&1.3507   &1.2020 \\
			\multicolumn{1}{c|}{$QS_{1}$ $ \left\langle {{x}_{\text{B}}} \right\rangle ,\left\langle {{p}_{\text{B}}} \right\rangle $} &0.2737  &0.2741 \\
			\multicolumn{1}{c|}{$QS_{1}$ $ \left\langle x_{\text{B}}^{2} \right\rangle ,\left\langle p_{\text{B}}^{2} \right\rangle  $ }&1.2769  &1.2757 \\
			\multicolumn{1}{c|}{$QS_{2}$ $ \left\langle {{x}_{\text{B}}} \right\rangle ,\left\langle {{p}_{\text{B}}} \right\rangle $}  &$-8.76\times {{10}^{-4}}$  &0.3872  \\
			\multicolumn{1}{c|}{$QS_{2}$ $ \left\langle x_{\text{B}}^{2} \right\rangle ,\left\langle p_{\text{B}}^{2} \right\rangle  $ }&1.2019   &1.3509 \\
			\multicolumn{1}{c|}{$QS_{3}$ $ \left\langle {{x}_{\text{B}}} \right\rangle ,\left\langle {{p}_{\text{B}}} \right\rangle $} &-0.2744   &0.2739 \\
			\multicolumn{1}{c|}{$QS_{3}$ $ \left\langle x_{\text{B}}^{2} \right\rangle ,\left\langle p_{\text{B}}^{2} \right\rangle  $ }&1.2756  &1.2756 \\
			\multicolumn{1}{c|}{$QS_{4}$ $ \left\langle {{x}_{\text{B}}} \right\rangle ,\left\langle {{p}_{\text{B}}} \right\rangle $} &-0.3876  &$-5.87\times {{10}^{-4}}$  \\
			\multicolumn{1}{c|}{$QS_{4}$ $ \left\langle x_{\text{B}}^{2} \right\rangle ,\left\langle p_{\text{B}}^{2} \right\rangle  $ }&1.3518  &1.2012  \\
			\multicolumn{1}{c|}{$QS_{5}$ $ \left\langle {{x}_{\text{B}}} \right\rangle ,\left\langle {{p}_{\text{B}}} \right\rangle $} &-0.2745  &-0.2735  \\
			\multicolumn{1}{c|}{$QS_{5}$ $ \left\langle x_{\text{B}}^{2} \right\rangle ,\left\langle p_{\text{B}}^{2} \right\rangle  $ }&1.2757  &1.2761 \\
			\multicolumn{1}{c|}{$QS_{6}$ $ \left\langle {{x}_{\text{B}}} \right\rangle ,\left\langle {{p}_{\text{B}}} \right\rangle $} &$-6.26\times {{10}^{-4}}$   &-0.3877 \\
			\multicolumn{1}{c|}{$QS_{6}$ $ \left\langle x_{\text{B}}^{2} \right\rangle ,\left\langle p_{\text{B}}^{2} \right\rangle  $ }&1.2008   &1.3517 \\
			\multicolumn{1}{c|}{$QS_{7}$ $ \left\langle {{x}_{\text{B}}} \right\rangle ,\left\langle {{p}_{\text{B}}} \right\rangle $} &0.2731   &-0.2741 \\
			\multicolumn{1}{c|}{$QS_{7}$ $ \left\langle x_{\text{B}}^{2} \right\rangle ,\left\langle p_{\text{B}}^{2} \right\rangle  $ }&1.2759   &1.2760 \\
			\hline  
	\end{tabular}}
\end{table}
\begin{figure}[!ht] 
	\centering\includegraphics[width=8.8cm]{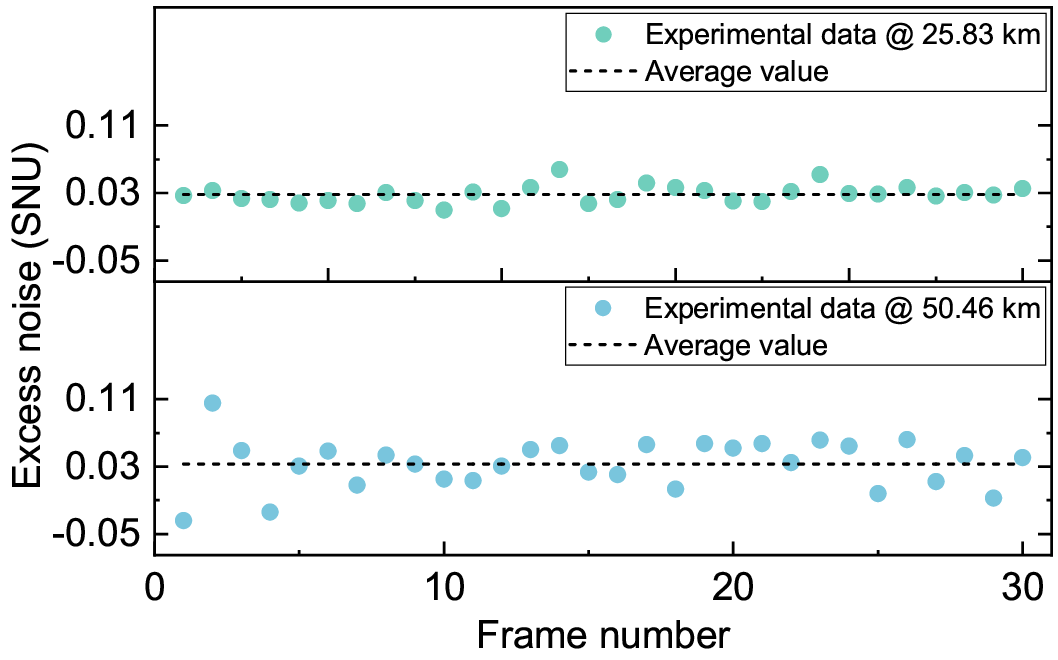}\\
	\caption{\label{The_excess_noise} The excess noise of the integrated CV-QKD system at different transmission distances. The green and blue solid circles represent the experimental excess noise at 25.8 and 50.4 km, respectively. The black dashed line represents the average value of the experimental points.}
\end{figure}
\subsubsection{Receiver}
In general, It is difficult to have a flat response spectrum within the bandwidth of a realistic high speed heterodyne detector. Similar to the high-frequency roll-off of the transmitter, the uneven response spectrum causes distortion of the quantum signal and crosstalk between adjacent quantum signals, which introduce measurement noise to the quantum signal. We designed a digital post-equalizer (whitening filter) to handle the acquired electronic noise, shot noise, and the quantum signals. The equalizer flattens the spectral response of the heterodyne detector based on the inverse frequency response of the directly measured shot noise of the heterodyne detector, as shown in Fig. \ref{DSP}(b). The whitening filter’s cutoff frequency is set at 2.71 GHz to remove the measured residual carrier at 2.78 GHz. Based on the estimated frequency difference between the pilot tone and the LO, the quantum signal is down-converted to the baseband and low-pass filtered using a RRC matched filter, and then downsampled to 1.5625 GBaud. The pilot tone is transformed to the baseband and filtered through a 100 kHz low-pass filter to estimate the relative phase between the LO and the quantum signal. Next, the time delay caused by the quantum signal transmission through the quantum channel is estimated based on the cross correlation between Bob’s measured data and Alice’s modulation data. The fast phase fluctuations of the quantum signals were corrected by the estimated phase from the pilot tone and the slow residual phase drift was compensated by rotating the quantum signals in a block manner. More precisely, the optimal demodulation phase of the quantum signal is searched by maximizing the correlation between Alice’s and Bob’s data.\\
\indent Because both the quantum signal and the pilot tone are generated by modulating the same optical carrier, their relative phase remains constant. Moreover, since they propagate along the same optical path and considering that the frequency difference between the quantum signal and the pilot tone is very small (therefore the dispersion effect can be neglected), their phase variations due to the optical path fluctuations are consistent. Therefore, the pilot tone can be used to estimate the relative phase between the LO and quantum signal.\\ 
\indent To accurately estimate the frequency deviation and phase fluctuation, the pilot tone’s signal-to-noise ratio after filtering is set to approximately 40 dB. The measured phase drift before the phase recovery is shown in Fig. \ref{The phase drift}, the results indicate that the phase recovery accuracy is within approximately 0.2°. The measured phase shift is around 10 krad/s, which consist of the rapid phase drift due to the independent lasers and the phase fluctuations arising from the long-distance single mode fiber.
\subsection{Experimental Results}
The experimental parameters for our integrated QKD system are listed in Table \ref{tab:table}. Table \ref{tab:table2} shows the experimentally measured first and second moments. The modulation variance of the quantum signal $V_\text{A}$ is selected by considering both the optimal modulation variance and induced excess noise (the observed excess noise is proportional to the modulation variance). The heterodyne detector operates at a shot noise to electrical noise ratio of 7 dB, under this condition the detector exhibits good stability and linearity that can minimizing the excess noises introduced by the measurement process.\\
\indent Low detection efficiency (0.37) is a key factor affecting the system performance. In the experiment, we calibrate the detection efficiency of the silicon photonic integrated receiver by using the same heterodyne detector with a pair of fiber-coupled photodiodes (with known quantum efficiency) as a benchmark.\\
\begin{figure}[!ht] 
	\centering\includegraphics[width=8.8cm]{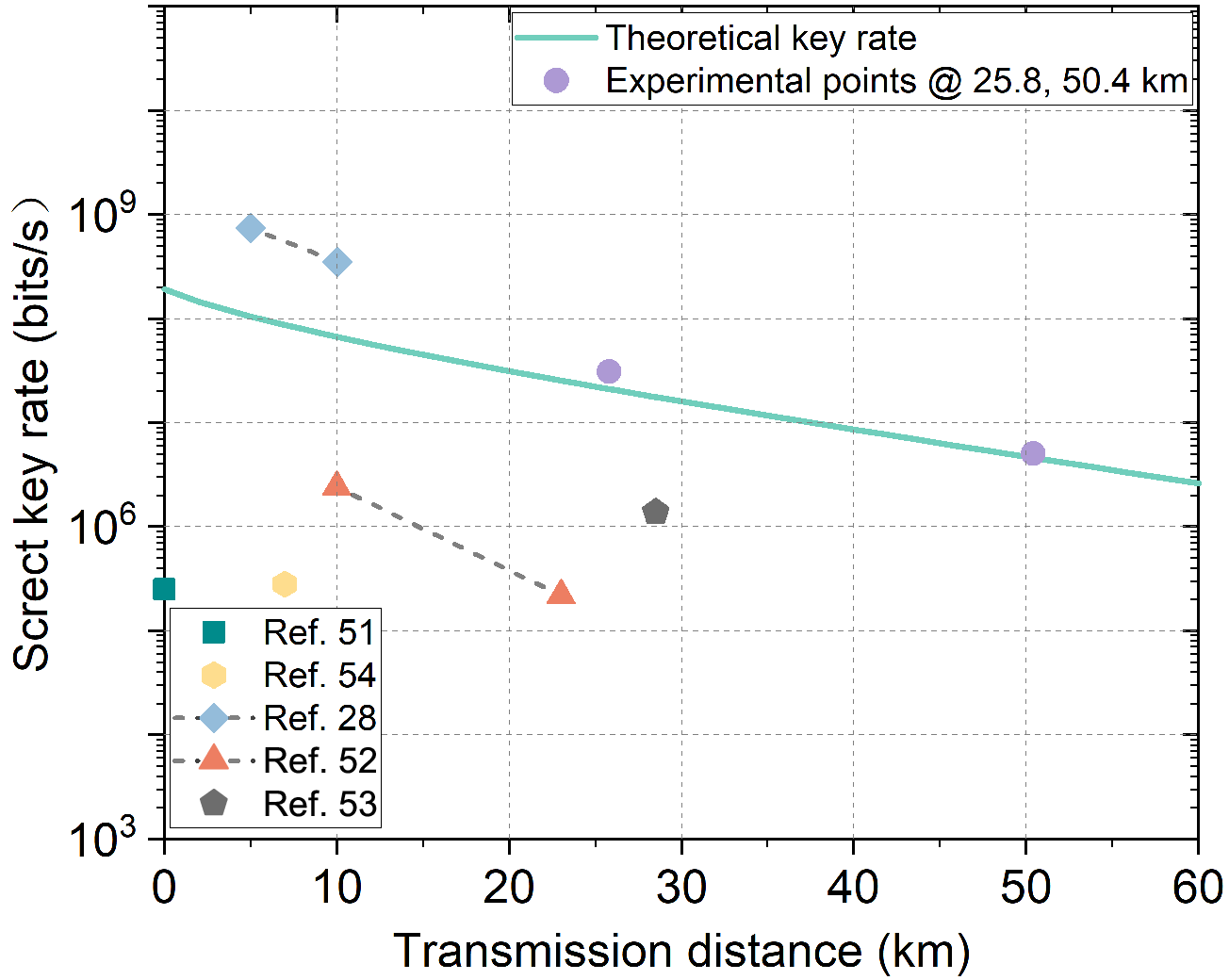}\\
	\caption{\label{Rate} The secret key rates of the integrated CV-QKD system. The green curve shows the theoretical secret key rate based on the experimental parameters, and the purple circles correspond to the experimental results. The green squares \cite{41}, yellow hexagon\cite{44}, blue diamonds \cite{25}, red triangles \cite{42}, and black pentagons \cite{43} represent the results of current integrated CV-QKD systems in literatures.}
\end{figure} 
\indent To characterize the stability of our system, 30 frames of data were recorded and the system’s excess noise are estimated under different transmission distances, as shown in Fig. \ref{The_excess_noise}. The data size for each frame are $8\times10^7$ and $1.28\times10^8$ and the corresponding average excess noise are 0.028 and 0.033 SNU at transmission distance of 25.8 and 50.4 km, respectively. The results indicate that the excess noise of our integrated CV-QKD system mainly originates from the preparation noises of the quantum states, and the remaining excess noise comes from the measurement noises and quantum channel. As the transmission distance increases, the fluctuation of the excess noise becomes significant. This is due to that the excess noise is referred to the quantum channel input, therefore the statistical fluctuations effect of the measured excess noise at Bob’s station will be amplified by a factor equal to the reciprocal of the total efficiency of the quantum channel transmission and detection.\\
\indent Figure \ref{Rate} illustrates the secret key rates of the integrated CV-QKD system in the asymptotic case. The purple circles represent the experimentally asymptotic secret key rates at 25.8 and 50.4 km. The green curve denotes the theoretical predictions using the experimental parameters where the modulation variance and the excess noise takes the mean value of the results at 25.8 and 50.4 km. The results of previous integrated CV-QKD were also plotted for comparison. Except for Ref. \cite{41} where both the transmitter and receiver are on-chip integrated, other works \cite{25,42,43,44} focus on the demonstration of the integrated receiver. Our integrated CV-QKD system achieves secret key rate of 31.05 and 5.05 Mbps over 25.8 and 50.4 km standard single-mode fiber, respectively, which is the longest distance so far for integrated CV-QKD system, to our knowledge.
\section{Conclusion}
\indent We present a high-performance integrated CV-QKD system with 8-PSK discrete modulation based on integrated silicon photonic transmitter and receiver. The system operates with a symbol rate of 1.5625 GBaud and achieves highly asymptotic secret key rates of 31.05 and 5.05 Mbps over 25.8 and 50.4 km standard single-mode fiber, respectively. Our results pay the way for the quantum secure communication network at metropolitan area on a large-scale. In the future, we will integrate the laser source (for example, a III–V InP laser diode) on the silicon photonic transmitter with a hybrid integration technique \cite{50}. In the receiver side, the coupling loss that directly relate to the detection efficiency could be decreased to 1.2 dB level by using advanced packaging technique \cite{37}. Furthermore, the symbol rate of the integrated system could be improved by high speed, low noise circuit design and production. A silicon photonic dynamic polarization controller could also be incorporated in the receiver for polarization tracking \cite{51}.

\providecommand{\noopsort}[1]{}\providecommand{\singleletter}[1]{#1}%


\end{document}